\documentclass{ws-ijmpa}

\begin{document}

\markboth{Swanson}
{Review of Heavy Hadron Spectroscopy}

%
\catchline{}{}{}{}{}
%

\title{Review of Heavy Hadron Spectroscopy}

\author{\footnotesize Eric Swanson}

\address{Department of Physics and Astronomy, University of Pittsburgh\\
Pittsburgh, Pennsylvania, 15260\\
United States of America}

\maketitle


\begin{abstract}
The status of some of the recently discovered heavy hadrons is presented.

\end{abstract}

\section{Introduction}
More than 30 years after the November Revolution, charmonium spectroscopy continues to 
surprise and challenge. A new era began in April of 2003 when BaBar announced the
discovery of the enigmatic $D_s(2317)$. This state continues to perplex us and 
Fermilab, CLEO, BES, BaBar, and Belle continue to add grist to the mill. This conference
report is a  brief review of the new heavy hadron spectroscopy.

\section{The New States}

\subsection{$B_c$}

CDF recently announced the discovery of the $B_c$ meson. The timing was arranged with the
FNAL lattice group to permit them a prediction of its mass. The respective results are
6287(5)(1) MeV\cite{cdfA} and 6304(4) MeV\cite{kronfeld}.  Thus CDF may claim a state, the lattice 
may claim a victory, and the $B_c$ appears to carry no surprises. Historians may be interested to
know, however, that Godfrey and Isgur\cite{GI} predicted this mass with the same accuracy as the
lattice, but with a twenty year lead time.

\subsection{$h_c$}

The $h_c$ has been observed by CLEO\cite{cleoB} with a mass of 3524(1) MeV.
This may be compared with typical quark model expectations\cite{BGS} of 3518 MeV, leading
one to suspect that the $h_c$ is also not hiding anything from us. Nevertheless, it tells us
something: the $h_c$ lies within 2 MeV of the spin-averaged mass of the $\chi_{cJ}$ multiplet:
$M_{c.o.g.} = {1\over 9}(\chi_0 + 5 \chi_1 + 9 \chi_2) = 3525.36$. Since the splittings between
these states are driven by the ${\cal O}(v/c)$ quark interactions, we
are learning something about the Dirac structure of confinement (namely that it is an effective
scalar).

\subsection{$\eta_c'$}

The CLEO collaboration\cite{cleoC} has observed the $\eta_c'$ 
with a mass of 3638(4) MeV and a width of 19(10) MeV. This is to be compared with the 
old result of Crystal Ball\cite{cball} of 3592(5) MeV. 
The $\eta_c'$ is of some interest because its splitting with the $\psi'$ is driven by
the hyperfine interaction and hence probes this interaction in a new region. Specifically, the ground
state vector-pseudoscalar splitting is
$m(J/\psi) - m(\eta_c) = 117$ MeV
whereas the excited splitting is now measured to be
$m(\psi') - m(\eta_c') = 48$ MeV.

Theoretical expectations for the former range from 108 to 123 MeV in simple (or `relativised') 
quark models\cite{BGS} and thus are within expectations. 
Alternatively, the latter is predicted to be 67 MeV in the quark model of Eichten,
 Lane, and Quigg\cite{ELQ}. However, the authors note that including
unquenching effects due to open charm meson loops lowers this splitting to 46 MeV  which is
taken as evidence in favour of their `unquenched' quark model. However, the simple quark models 
mentioned above\cite{BGS} find splittings of 42 - 53 MeV,
indicating that it is too early to make definitive conclusions about loop effects.
It is worth noting, furthermore, that attempts to unquench the 
quark model are fraught with technical difficulty\cite{screen} and a great deal of
effort is required before we can be confident in the results of any model.

\subsection{$D_s(2317)$ and $D_s(2460)$}

These states are roughly 100 MeV below quark model expectations and point to 
either exotic structure, such as $DK$ molecules\cite{bcl}, or to a deep misunderstanding
of heavy-light hadrons. For example, the $DK$ and $D^*K$ continua are both nearby and 
couple to $D_{s0}$ and $D_{s1}$ in S-wave. Is it possible that we have underestimated
the importance of coupled channel effects in some systems?


\subsection{$D_s(2632)$}

The $D_s(2632)$ was discovered by the SELEX collaboration at FNAL\cite{selex} in the
final states
$D^0K^+$ and $D_s \eta$. The measured mass is 2632.6(1.6) MeV and the state is
surprisingly narrow with a width of
less than 17 MeV at the 90\% confidence level. 
The ratio of the partial widths is measured to be 

\begin{equation}
{\Gamma(D_s \to D^0K^+) \over \Gamma(D_s \to D_s\eta)} = 0.16 \pm 0.06.
\end{equation}
As pointed out by the SELEX collaboration, this is an unusual result since the $DK$ mode is favoured
by phase space. 

It is unlikely that this state is a $c\bar s$ hybrid since the mass of such a state is expected to
be roughly 3170 MeV. Possible molecular states include a $D_s^*\eta$ system at 2660 MeV or 
$D_s^* \omega$ or $D^*K^*$ states at 2900 MeV. However the former is a P-wave which is not favoured
for binding, while the latter are too heavy to be plausible. 

The remaining possibility is that the $D_s(2632)$ is a 
radially excited $c\bar s$ vector\cite{chao} (although it is some 100 MeV lighter than quark model predictions
of 2730 MeV). The peculiar decay ratio remains to be explained. Experience with the decay modes
of the $\psi(3S)$ to $DD$, $DD^*$ and $D^*D^*$ points to a possible resolution:
transition matrix elements for excited hadrons may have zeroes due to wavefunction nodes.
It is possible that such a node is suppressing the $DK$ decay mode.
Computation\cite{gang} reveals that there is indeed a node but that it occurs at a wavefunction scale which
is 20\% lower than preferred. Furthermore, the $DK$ mode is always larger than
the $D_s\eta$ mode. 

We have run out of options and must conclude that the $D_s(2632)$ is an experimental artefact.
This conclusion now appears likely because  searches by FOCUS\cite{focus},
BaBar\cite{babarA}, and CLEO\cite{cleoD} have found no evidence for the state.

\subsection{$X(3872)$}

The $X(3872)$ is the poster boy of the new heavy hadrons -- it has been confirmed by four 
experiments\cite{Xexpts} at a mass of 3872 MeV and is very narrow, $\Gamma < 2.3$ MeV at 95\%.
The anomalous nature of the $X$ has led to much speculation: tetraquark\cite{maianiA},  
cusp\cite{bugg}, hybrid\cite{li}, or glueball\cite{seth}. But the most popular explanation
is that it is a $D\bar D^*$ bound state\cite{Xmol,ess}. This model has successfully 
predicted\cite{ess} the
quantum numbers of the $X$ ($J^{PC} = 1^{++}$\cite{olsen}), the decay mode $\pi\pi\pi J/\psi$\cite{belleE}, 
that the
$\pi\pi$ and $3\pi$ modes should be comparable\cite{belleF}, and that the $\pi\pi$ invariant 
mass distribution should be
dominated by the $\rho$\cite{cdf} while the $3\pi$ invariant mass distribution should be dominated
by the $\omega$\cite{belleE}.

This string of successes has met with recent experimental challenges:
(i)
The $X$ has been observed decaying to $\gamma J/\psi$\cite{belleE} with a strength $Br(X\to \gamma J/\psi) /
Br(X \to \pi\pi J/\psi) = 0.14 (5)$\cite{belleH}. This rate is substantially larger than predicted in 
the model of Ref. \cite{ess}. (ii) There are rumours that Belle have seen the mode $X \to D\bar D \pi$ and
that its rate is ten times larger than that of $\pi\pi J/\psi$\cite{DDpi}. The model predicts this ratio
to be 1/20.  (iii) BaBar report\cite{denis} 

\begin{equation}
{Br(B^0 \to X K^0)\over Br(B^+ \to X K^+)} = 0.61(36)(6).
\end{equation}
This is at odds with the molecular picture which predicts a ratio of ${\cal O}({1\over N_c^2}) + {\cal O}({Z_{D^+D^-}\over Z_{D^0\bar D^{0*}}}) \approx 10 \%$.
(iv) Belle\cite{olsen} and BaBar\cite{denis} have measured the product of branching ratios:

\begin{equation}
Br(B^+ \to X K^+) Br(X \to \pi\pi J/\psi) = 1.3(3) [0.85(30)] \cdot 10^{-5}.
\end{equation}
The new $D\bar D \pi$ data imply that $Br(B \to \pi\pi J/\psi) < 0.1$ which implies that $Br(B \to XK) > 10^{-4}$. This is comparable to the rate $Br(B \to \chi_{c1}K)$ and points to a large $c\bar c$ component
in the $X$.

All of these new data may be accounted for if the predicted hidden charm interactions of Ref.\cite{ess}
were over-estimated. This leads to weaker binding which gives rise to a much narrower $X$ with
a dominant $D\bar D\pi$ mode, weak $\pi\pi J/\psi$ and $\pi\pi\pi J/\psi$ modes, and a radiative transition
of the desired magnitude.

%

\subsection{$X(3940)$}

The $X(3940)$ is seen by Belle\cite{belleC} recoiling against $J/\psi$ in $e^+e^-$ collisions. The
state has a mass of 3943(11)(13) MeV and a width of 87(22)(26) MeV\cite{trabelsi}. The $X$ is
seen to decay to $D\bar D^*$ and not to $\omega J/\psi$ or $D\bar D$.
It is natural to attempt a $2P$ $c\bar c$ assignment for this state since the expected mass 
of the $2^3P_J$ multiplet is 3920 - 3980 MeV and the expected widths are 30 - 165 MeV\cite{BGS}.
Finally, if the $D\bar D^*$ mode is dominant it suggests that the $X(3940)$ is the $\chi_{c1}'$.
The problem  with this assignment is that there is no evidence for the $\chi_{c1}$ in the
same data.
This has led Olsen to
speculate\cite{Olsen2} that the $X$ is the $\eta_c''$. Unfortunately this interpretation is also
suspect because the $\eta_c''$ has an expected mass of 4064 MeV, 120 MeV too high.

\subsection{$Y(3940)$}

The $Y(3940)$ is claimed as a resonance in the $\omega J/\psi$ subsystem of the decay
$B \to K\pi\pi\pi J/\psi$\cite{belleB} with a mass of 3940(11) MeV and a width of 92(24)
MeV. The state has not been seen in the decay modes $Y \to D\bar D$ or $D\bar D^*$.
Again, the mass and width of the $Y$ suggest a radially excited P-wave charmonium. However,
the $\omega J/\psi$ decay mode is peculiar.
In more detail, Belle measure $Br(B \to KY) Br(Y \to \omega J/\psi) = 5.0(9)(16)\cdot 10^{-5}$.
One expects that $Br(B \to K \chi_{cJ}') < Br(B\to K\chi_{cJ}) = 4(1)\cdot 10^{-4}$.
This implies $Br(Y \to \omega J/\psi) > 12 \%$, which is unusual for a canonical $c\bar c $ state
above open charm threshold.
 
Thus the $Y$ is something of an enigma, driving the claim of the Belle collaboration that it
is a hybrid.  This is perhaps premature -- certainly more data are required before strong claims
can be made.

\subsection{$Z(3930)$}

This state was observed  by the Belle collaboration\cite{belle} in $\gamma\gamma \to D\bar D$ with a mass of 3931(4) MeV and a width of 20(8)(3) and a claimed significance of 5.5 sigma. 
The $D\bar D$ helicity distribution is consistent with J=2. In line with the $X$ and the $Y$, the
$Z$ seems an
obvious candidate for the $\chi_{c2}'$ (the $\chi_{c1}'$ cannot decay to $D\bar D$). The predicted 
mass of the $\chi_{c2}$ is 3972 MeV and the predicted width is 80 MeV\cite{BGS}. However,
setting the mass to the measured 3931 MeV restricts phase space sufficiently that the
predicted strong width drops to 47 MeV, reasonably close to the measurement. The 
predicted branching fraction to $D\bar D$ is 70\%. The largest radiative transition is 
$\chi_{c2}' \to \psi' \gamma$ with a rate of 180(30) keV. At this stage we have no reason not
to believe that the $Z$ is the previously unknown $\chi_{c2}'$.

\subsection{$Y(4260)$}

The $Y(4260)$ was discovered as an enhancement in the $\pi\pi J/\psi$ subsystem of the reaction
$e^+e^- \to \gamma_{\rm ISR} \psi \pi\pi$ with a mass of 4259(8)(4) MeV and a width
of 88(23)(5) MeV  
by the BaBar collaboration\cite{babarB}.  Evidentally the state is a vector
with $c \bar c$ flavour. Of course the low lying charmonium vectors are well known and the only
mesonic charmonium vector available is the $\psi(3D)$. However quark model estimates of its mass
place it at 4460 MeV, much too heavy for the $Y$. Of course this statement relies on the quark model
itself -- Llanes-Estrada\cite{filipe} has argued that the $Y$ is the $\psi(4S)$ based on 
the spectrum
of a relativistic model. Maiani {\it al.}\cite{maiani} claim the $Y$ is a tetraquark $c\bar c s\bar s$
state which decays predominantly to $D_s\bar D_s$. 
Of the states which we know must exist, the most natural explanation is as a 
$c\bar c$ hybrid\cite{Yhybrid}. The
lightest charmonium hybrid is expected at 4400 MeV, somewhat high, but perhaps acceptable
given our lack of experience in this sector.

%


Lastly, it is tempting to examine molecular interpretations of this state. In particular
$DD_1$ is an S-wave threshold at 4290 MeV -- close enough that the enhancement may simply be
a cusp effect. If the system does bind, it does so with a novel mechanism since pion exchange
does not lead to a diagonal interaction in this channel (unlike the case of the $X(3872)$).
Off-diagonal interactions may provide the required novelty.



\section{Conclusions}

The new heavy meson spectroscopy is no mere butterfly collecting --  
the $D_s$ spectrum and the $X$s, $Y$s, and $Z$s challenge our understanding of QCD. Can we
rise to the challenge?  It is clear that the simple constituent quark model must fail somewhere
(gluonic degrees of freedom turn on, relativity and chirality become important, and coupled channels
become dense) -- are we seeing this? Lastly, have we entered a new era of `mesonic nuclear physics'?

\section*{Acknowledgments}
This work was supported by PPARC grant PP/B500607 and the US Department of
Energy under contract DE-FG02-00ER41135.
I am grateful to Ted Barnes, Denis Bernard, Frank Close, Steve Godfrey, and Steve Olsen for
many interesting discussions on the new states. I wish to thank the organisers and especially
Xiangdong Ji for the invitation to discuss this fascinating topic in Beijing.

\end{document}